\documentclass[aps,prb,showpacs]{revtex4}

\usepackage{amsfonts,amsmath,amssymb}
\usepackage{graphicx,color}
\usepackage{verbatim}

\begin{document}

\title{Simulation of structural phase transitions in NiTi}

\author{Daniel Mutter}
\email{daniel.mutter@uni-konstanz.de}
\affiliation{Department of Physics, University of Konstanz, 78457 Konstanz, Germany}
\author{Peter Nielaba}
\affiliation{Department of Physics, University of Konstanz, 78457 Konstanz, Germany}

\date{\today}

\begin{abstract}
By means of molecular-dynamics simulations, temperature driven diffusionless structural phase transitions in equi- and nearly equiatomic ordered nickel-titanium alloys were investigated. For this purpose, a model potential from the literature was adopted [W.~S.~Lai and B.~X.~Liu, J.~Phys. Condens. Matter 12, L53 (2000)], which is based on the tight-binding model in second moment approximation. The model predicts a stable B19$'$ phase at low temperatures and a nearly cubic B2 phase at high temperatures. After an analysis of crystallography and energetics of the emerging structures, the experimentally known strong dependence of transition temperatures on composition is confirmed and related to lattice instability. Free energy calculations finally give insight into the driving forces of the phase transitions, and reveal free energy barriers inhibiting them below the transition temperatures.
\end{abstract}


\maketitle

\section{Introduction}\label{sec1}

The possibility to recover its original shape after deformation and heating [``shape-memory effect'' (SME)] as well as the effect of superelasticity (SE), where elastic strains of up to 8\% can be achieved \cite{buchots:98}, make shape memory alloys (SMA) to widely used functional materials in industry, ranging from dental to aerospace applications. The physical origin of SME and SE is a diffusionless structural phase transition between a high-temperature phase (``austenite'') and a low-temperature phase (``martensite'') with lower symmetry. In order to explore this kind of phase transitions theoretically at the nanoscale, molecular dynamics simulations have been performed in the past for different alloys such as NiAl by Saitoh and Liu \cite{sailiu:09} or FeNi by Entel and coworkers \cite{entkad:98, entmey:00}.\\
\indent Among all, the most commonly used SMA's in industry are nickel-titanium alloys with equi- or near-equiatomic composition. In these alloys, the experimentally observed ground state structure consists of monoclinic lattice cells with additional in-plane and out-of-plane shuffles of the atoms at the faces of the cell \cite{otsren:99}. This ordered, bi-atomic alloy structure is referred to as B19$'$ and has space group $P2_1/m$. If the crystal is heated above a strongly concentration dependent temperature \cite{hanbut:67}, it undergoes a structural change to the more symmetric, body-centered cubic B2-phase. In addition, the so-called R-structure, an intermediate phase with trigonal symmetry, appears during the martensitic B2$\rightarrow$B19$'$-transition in some experiments \cite{sitsch:03, zuwan:04}, but there is no overall agreement among them concerning the correct space group. By means of \emph{ab initio} calculations, the phase energetics and lattice parameters of the proposed structures have been analyzed in many publications \cite{pascol:95, biheib:96, sanalb:98, parpar:02, huaack:03, luhu:07, kibseh:09, hatkon:09.2, visstr:10}. Among those, Huang \emph{et al.} \cite{huaack:03} were the first, who proposed the body-centered orthorhombic B33 to be the most stable martensitic ground state of NiTi, but this was not yet observed in experiments.\\
\indent This ambiguity of the exact crystal structures on the one hand and discrepancies in the predictions of elastic constants on the other hand \cite{ackjon:08} make it difficult to construct an interatomic potential suitable for dealing with all the observed phenomena in NiTi, especially the structural transitions with the associated shape memory and superelastic behavior, by means of molecular dynamics (MD) simulations. That is why only a few potentials exist in the literature, but of those, each is capable of describing some aspects of the material properly: Farkas \emph{et al.} \cite{farroq:96} proposed a potential based on the embedded-atom method (EAM), which is able to reproduce lattice parameters and cohesive energies of the compounds B2 NiTi and Ni$_3$Ti, but fails in stabilizing a monoclinic structure in the ground state. With a potential originating from the second-moment approximation of the tight-binding model (TBSMA), Lai and Liu \cite{lailiu:00} studied crystalline-to-amorphous transitions of nickel/titanium solid solutions as well as the amorphization of several NiTi compounds upon ion irradiation. Since this potential predicts a monoclinic structure to have a higher cohesive energy than the cubic state, it was adopted by Sato \emph{et al.} \cite{satsai:06} to perform MD on the stress induced martensitic phase transformation, whereby multiple B2-B19$'$ pathways could be identified. Until today, there is a lack of studies regarding the temperature driven structural changes in NiTi with simulations. Only Ishida and Hiwatari \cite{ishhiw:07} report on simulations using the modified EAM, where a reversible phase transition at a certain temperature occurs, but the crystallographies of the parent and martensitic phases are not satisfactorily clarified.\\
\indent In the present work, the applicability of the TBSMA potential for performing reliable MD simulations of this kind of transformation in NiTi is analyzed. For that purpose, the potential is extended by a function, which controls the cutoff-behavior of the rapidly decreasing exponential functions appearing in this model, which describe the hopping integrals and the pair interaction. This is explained in Sec.\:\ref{sec2A}, together with some simulation details. In Sec.\:\ref{sec2B}, a method for calculating the free energy in a MD simulation is described, by which the thermodynamics of the phase transition can be studied. The modified TBSMA potential leads to a better agreement of lattice parameters and energetics of the stable B19$'$-structure with experiments and \emph{ab initio} calculations, which is covered in Sec.\:\ref{sec3A}. In addition, MD simulations of equiatomic NiTi, performed under periodic boundary conditions and at varying temperatures are presented there, in which first order structural phase transformations can be identified, and the emerging structures are analyzed. In Sec.\:\ref{sec3B}, the dependence of transition temperatures on Ni/Ti concentration is considered and compared with experiments. The strong decrease of these temperatures when the Ni content differs slightly from 50\% is shown to be attended by a considerable destabilization of the lattice structure. Thermodynamic free energy and entropy calculations during the heating and cooling process of the system are presented in Sec.\:\ref{sec3C}. In addition, free energy barriers and their dependence on temperature and concentration are calculated along a linear transformation path. Finally, a summary and conclusion are given in Sec.\:\ref{sec4}.

\section{Theory}\label{sec2}

\subsection{Interatomic potential and simulation method}\label{sec2A}

The semi-empirical potential applied in this work to perform MD on the martensitic phase transitions in NiTi originates from the tight-binding-bond model in second-moment approximation as described by Cleri and Rosato \cite{cleros:93}. According to this, the total energy $U_t$ of a system is written as a sum over all atoms $i$:
\begin{equation}
U_t = \sum_i\left(U_i^B+U_i^R\right),
\label{g0}
\end{equation}
with
\begin{equation}
U_i^B = -\sqrt{\sum_{j\neq i}\xi^2_{\alpha\beta}\exp\left[-2q_{\alpha\beta}\left(\frac{r_{ij}}{d_{\alpha\beta}}-1\right)\right]}
\nonumber
\end{equation}
and
\begin{equation}
U_i^R = \sum_{j\neq i}A_{\alpha\beta}\exp\left[-p_{\alpha\beta}\left(\frac{r_{ij}}{d_{\alpha\beta}}-1\right)\right].
\nonumber
\end{equation}
The quantum mechanical many-body character of metallic bonding is expressed in the bond energy $U_i^B$, which is approximated as the negative square root of the second moment of the electron density of states at atom $i$. This is given as a sum over squared two-center- (``hopping''-) integrals, with an effective part $\xi_{\alpha\beta}$ and an exponential distance dependence. $U_i^R$ represents a pairwise Born-Mayer repulsion \cite{bormay:32}, which is necessary to stabilize the crystal. $r_{ij}$ denotes the distance between atoms $i$ and $j$, and the indices $\alpha\beta$ account for different atomic types, what leads to 15 parameters in binary alloys AB ($\xi,\ q,\ A,\ p,\ d$ for A-A, B-B and A-B interaction).\\
\indent In the case of NiTi, they were determined by Lai and Liu \cite{lailiu:00} by calculating properties like cohesive energies, lattice parameters and elastic constants analytically from the potential and fitting them to \emph{ab initio} data of the pure materials and of the B2-NiTi phase at $T$\,=\,$0$~K. Additionally, they treated the cutoff radius of the exponential functions as an adjustable quantity, and proposed $r_c$\,=\,$4.2~\mathring{\mbox{A}}$ as an optimal choice. With this value, first and second nearest neighbors in NiTi are taken into account in the calculation, but since it lies closely below the distance of the third nearest neighbor shell, a smooth cutoff behavior has to be ensured in order to perform reliable MD simulations without diverging forces. Therefore, the cutoff function $f_c$ proposed by Baskes \emph{et al.} \cite{basnel:89}:
\begin{equation}
f_c = \left\{\begin{array}{l}
1,\ r\le r_c-\delta\\
\frac{1}{2}\left(1+x\right)-\frac{5}{8}x\left(x^2-1\right)+\frac{3}{16}x\left(x^4-1\right),\ r_c-\delta<r<r_c\\
0,\ r\ge r_c
\end{array}
\right.
\nonumber
\end{equation}
is multiplied to the exponential functions in the present work, with \mbox{$x=\left(r_c-\delta/2-r\right)/\left(\delta/2\right)$}. Good agreement with structural properties of the monoclinic B19$'$ ground state of NiTi is achieved by setting the additional parameter $\delta$ to 0.2~$\mathring{\mbox{A}}$.\\
\indent With this potential, MD simulations are performed with a velocity-Verlet algorithm \cite{swoand:82} (timestep $\Delta t$\,=\,$10^{-15}$ s) at different temperatures, imposed by a Nos\'{e}-Hoover thermostat \cite{nos:84}, and under periodic boundary conditions. A fully flexible simulation box is applied by the Parrinello-Raman method \cite{parrah:81} with improvements of Martyna \emph{et al.} \cite{martob:94}, and the pressure is set to 0~Pa.

\subsection{Free energy calculation}\label{sec2B}

The calculation of thermodynamic quantities in MD simulations can give deeper insight into phase behavior and phase transitions of many particle systems. While the internal energy is calculated straightforward by taking averages of the total energy $U_t$ (Eq.\:\ref{g0}), the free energy $F$ has to be obtained in a more complex way, e.g. by thermodynamic integration between a reference system with known $F$ and the system under consideration. In the present work, the method described by Frenkel and Ladd \cite{frelad:84} is applied, which makes use of an Einstein crystal (EC) as reference state, i.e. a system, where particles are bound to fixed lattice sites by harmonic springs. In the past, this method has successfully been used in combination with model potentials to describe e.g. bcc-fcc transitions in iron \cite{engsan:08}, liquid-crystal interfaces in Al \cite{meidav:92} or phase diagrams of Au-Ni alloys \cite{ogacar:02}.\\
\indent In detail, a parameter dependent potential $\tilde{U}\left(\lambda\right)$ is introduced, with $\tilde{U}\left(1\right)=U$, the potential of the system with unknown free energy, and $\tilde{U}\left(0\right)=U_{E}$, the potential of the EC:
\begin{equation}
\tilde{U}\left(\lambda\right) = U_E + \lambda\left(U-U_E\right).
\label{g1}
\end{equation}
It can easily be shown, that
\begin{equation}
\frac{\partial F\left(\lambda\right)}{\partial\lambda}=\langle\frac{\partial\tilde{U}\left(\lambda\right)}{\partial\lambda}\rangle_\lambda \nonumber
\end{equation}
with the canonical ensemble average $\langle\dots\rangle_\lambda$. With Eq.\:\ref{g1}, this leads to
\begin{equation}
F\left(\lambda=1\right) = F\left(\lambda=0\right) + \int_0^1\mathrm d\lambda\,\langle U-U_E\rangle_\lambda.
\label{g2}
\end{equation}
Since $F\left(\lambda=0\right)$, the free energy of the EC, is analytically known for a given temperature $T$, $F\left(\lambda=1\right)$ can be calculated in the simulations. Optimal spring constants for the EC are obtained by averaging and comparing the mean-squared displacements for $\lambda$\,=\,$0$ and $\lambda$\,=\,$1$. In the simulations presented here, this is done at first for 50000 timesteps at a given $T$, followed by a variation of $\lambda$ according to
\begin{equation}
\lambda:\ 1\xrightarrow{-0.05}0.1\xrightarrow{-0.005}0.01\xrightarrow{-0.0005}0, \nonumber
\end{equation}
since the integrand of Eq.\:\ref{g2} increases strongly when $\lambda$ approaches 0. At each $\lambda$-step, $\left(U-U_E\right)$ is averaged over 10000 timesteps, and at the end, a numerical integration yields the value of $F$. As described above, simulations are carried out at zero pressure, and therefore $F$ is equal to the Gibbs free energy $G$, which is used in the following, instead.

\section{Results and discussion}\label{sec3}

\subsection{Structures and Transformations}\label{sec3A}

\begin{figure}[!htbp]
\centering
\includegraphics[width=0.5\linewidth]{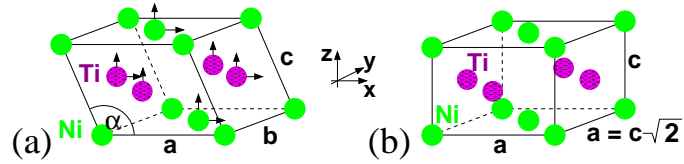}
\caption{(a) The monoclinic B19$'$ structure of NiTi with angle $\alpha$, lattice constants $a,\ b,\ c$, and shuffles, denoted by the arrows; (b) the cubic B2 structure.}
\label{f1}
\end{figure}

\begin{figure}[!htbp]
\centering
\includegraphics[width=0.5\linewidth]{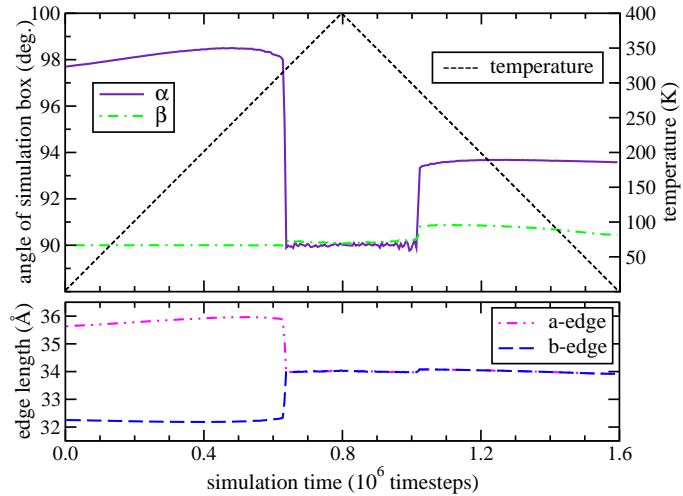}
\caption{Shape changes of the simulation box during a temperature cycle: (above) angles $\alpha$ (between $a$- and $c$-edge) and $\beta$ (between $a$- and $b$-edge) as well as the imposed temperature. (Below) lengths of $a$- and $b$-edges .}
\label{f2}
\end{figure}

\begin{figure}[!htbp]
\centering
\includegraphics[width=0.5\linewidth]{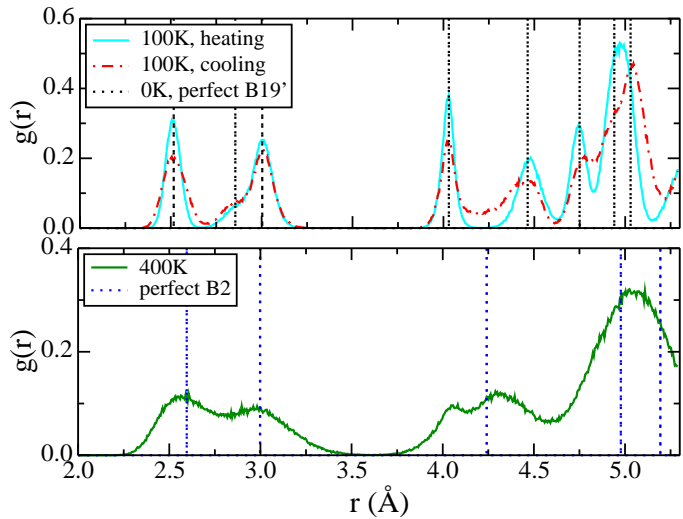}
\caption{Radial distribution function at different temperatures in the heating and cooling process together with the positions of the perfect structures.}
\label{f3}
\end{figure}

\begin{figure}[!htbp]
\centering
\includegraphics[width=0.8\linewidth]{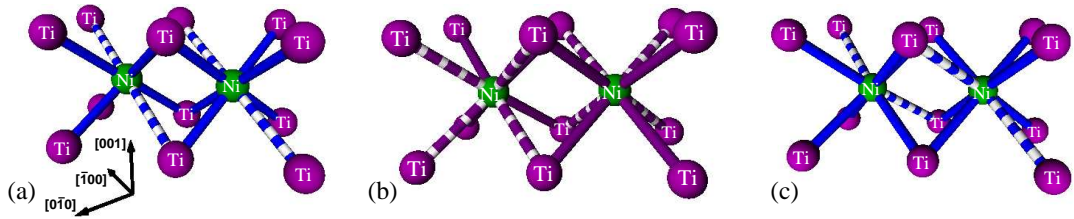}
\caption{Nearest neighbor (NN) environment of the emerging structures: (a) B19$'$ with NN distances of $2.51\ \mathring{\mbox{A}}$ (continuous) and $2.85\ \mathring{\mbox{A}}$ (dashed); (b) high-temperature B2 like phase with NN distances of $2.54\ \mathring{\mbox{A}}$ (continuous) and $2.69\ \mathring{\mbox{A}}$ (dashed); (c) ground state structure emerging upon cooling of the B2 like lattice, lengths as in (a).}
\label{f4}
\end{figure}

\begin{figure}[!htbp]
\centering
\includegraphics[width=0.5\linewidth]{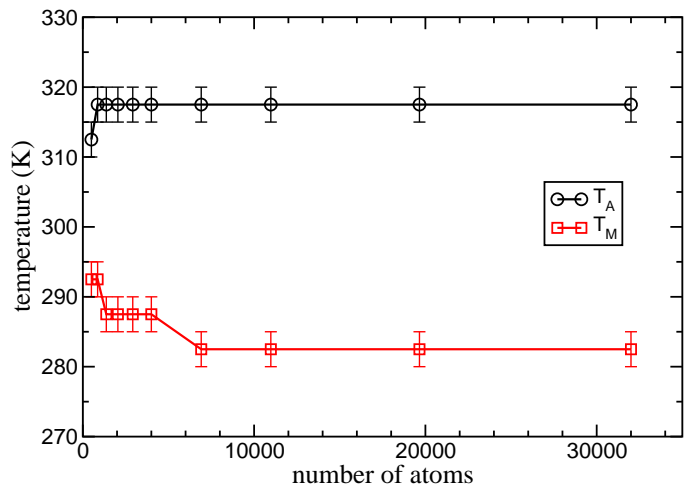}
\caption{Dependence of austenitic ($T_A$) and martensitic ($T_M$) transition temperatures on system size. The error bars result from a temperature step of 5 K during heating and cooling.}
\label{f10}
\end{figure}

In order to perform simulations of the temperature-driven structural phase transition between B19$'$ [Fig.\:\ref{f1}(a)] and B2 [Fig.\:\ref{f1}(b)] in NiTi, it has to be ensured at first, that the martensitic structure is stable at low temperatures. For this purpose, a system with 2048 particles and equiatomic composition was set up in a monoclinic structure with lattice parameters $a,\ b,\ c$ and $\alpha$ close to experimental values. Decreasing the temperature stepwise from $T$\,=\,$10$~K to $T$\,=\,$0$~K with a rate of 1~K per 10000 timesteps lead to a stable B19$'$-structure with parameters listed in Table 1.\\
\indent The comparison with experiments and \emph{ab initio} results shows deviations of about 2\%-5\% in the $a$, $b$ and $c$ parameters, and an excellent agreement concerning the value of the monoclinic angle $\alpha$, which is due to the modification of the TBSMA potential described above. The in-plane and out-of-plane shuffles with values between 2\% and 9\% of the cell parameters $b$ and $c$ do not appear within this model. A calculation of the cohesive energy leads to $E_c\left(\mbox{B19}'\right)$\,=\,$5.076$ eV/atom.
\\
\begin{table}[!htbp]
\caption{Calculated values of structural parameters (see Fig.\:\ref{f1}) and the cohesive energy difference to the B2 structure in comparison with results for the unmodified TBSMA potential, \emph{ab initio} calculations and experimental values. The lattice constants $a,\ b,\ c$ are given in $\mathring{\mbox{A}}$, the energies in meV/atom.}
\label{tab1}
\begin{tabular}[c]{l c c c c r}
\hline\hline
Structure&$a$&$b$&$c$&$\alpha$&$\Delta E_c$\\\hline
B2 {\small (this work)}&&&$3.01$&$90.0^{\circ}$&\\
B2$^a$&&&$3.01$&$90.0^{\circ}$&\\
B2$^b$&&&$3.019$&$90.0^{\circ}$&\\
B2$^c$&&&$3.013$&$90.0^{\circ}$&\\
B19$'$ {\small (this work)}&$4.45$&$4.03$&$3.00$&$97.7^{\circ}$&$54.0$\\
B19$'^a$&$4.46$&$4.19$&$2.96$&$93.3^{\circ}$&$18.0$\\
B19$'^b$&$4.677$&$4.077$&$2.917$&$98.0^{\circ}$&$55.42$\\
B19$'^d$&$4.66$&$4.11$&$2.91$&$98.0^{\circ}$&\\\hline\hline
\end{tabular}
\\{\flushleft
$^a$Calculated results with the unmodified tight-binding potential by Lai and Liu \cite{lailiu:00}.\\
$^b$Calculated \emph{ab initio} results by Hatcher \emph{et al.} \cite{hatkon:09.2}.\\
$^c$Experimental results by \v{S}ittner \emph{et al.} \cite{sitluk:03}.\\
$^d$Experimental results by Prokoshkin \emph{et al.} \cite{prokor:04}.\\}
\end{table}
\indent To check, whether the stable B19$'$ phase undergoes a structural phase transition to the B2 phase upon heating above a certain temperature, the system was heated about 1~K every 2000 simulation steps from 2~K to 400~K. In Fig.\:\ref{f2}, the behavior of the simulation box is recorded during this process.
\\
\indent The initially set angle $\alpha$ between the $a$- and $c$-edge of the simulation box [with respect to the cell shown in Fig.\:\ref{f1}(a)] first increases due to thermal expansion, but then drops immediately to $90^{\circ}$ when reaching 318~K, whereas the angle $\beta$ between the $a$- and $c$-edge stays nearly constant. At the same time, the different lengths of the $a$- and $b$-edges of the box reach the same value, which lies almost at the arithmetic mean. Since in addition neither the angle between the $b$- and $c$-edge ($90^{\circ}$) nor the length of the $c$-edge changes, a cubic box is obtained after the phase transition. Assuming a B2 structure, the lattice parameter $c$ can be calculated from the box dimensions, yielding $c$\,=\,$3.01\ \mathring{\mbox{A}}$. With this value, the cohesive energy of a perfect B2 lattice is computed within the model potential (Eq.\:\ref{g0}) to $E_c\left(\mbox{B2}\right)$\,=\,$5.022$ eV/atom, which leads to $E_c\left(\mbox{B19}'\right)-E_c\left(\mbox{B2}\right)$\,=\,$54$ meV/atom in good agreement with \emph{ab initio} results \cite{hatkon:09.2} (see Tab.\:\ref{tab1}). As the system is cooled down again, the box transforms with a hysteresis of 28~K to a shape, which differs from the starting B19$'$ geometry.\\
\indent To get more insight into the crystallography of the involved structures, the radial distribution function $g\left(r\right)$ is evaluated at different points in the temperature cycle, as shown in Fig.\:\ref{f3}. In the heating process at 100~K, where the phase transition has not occurred yet, the peaks of $g\left(r\right)$ lie at the positions of the perfect B19$'$ starting configuration, with the typical broadening due to thermal vibrations. Above the transition temperature, at 400~K, the structural change can be identified by the appearance of a peak at $r$\,=\,$4.3\ \mathring{\mbox{A}}$ and a lowering of the peaks at $r$\,=\,$4.47\ \mathring{\mbox{A}}$ and $r$\,=\,$4.75\ \mathring{\mbox{A}}$. The comparison of this structure with the previously assumed perfect B2 shows good agreement, only the peak at $r$\,=\,$4.04\ \mathring{\mbox{A}}$ indicates a slight difference. The transition occurring in the cooling process finally leads to a structure, for which the radial distribution function $g\left(r\right)$ shows nearly the same peaks as in the starting B19$'$ lattice, but since $g\left(r\right)$ cannot resolve the angular distribution of the appearing lengths, the identity of the structures cannot be concluded.
\\
\indent In order to clarify the crystallographic details, the nearest neighbor environment of the low- and high-temperature structures is analyzed in detail by averaging the neighbor distances of each atom over 10000 simulation steps. This is done at low temperatures before heating (B19$'$) and after cooling ($\overline{\mbox{B19}'}$), and at $T$\,=\,$350$ K (B2). The results for two characteristic atoms with nearest neighbors are shown in Fig.\:\ref{f4}.\\
\indent In the stable starting configuration B19$'$, two of the eight nearest neighbor (NN) distances are elongated about 13.5\% [dashed lines in Fig.\ref{f4}(a)] relative to the others due to the monoclinic shear along $\left[\bar{1}00\right]$. In an ideal B2 structure with $c$\,=\,$3.01\ \mathring{\mbox{A}}$ all the NN lengths would have the same value of $2.61\ \mathring{\mbox{A}}$. In fact, the high-temperature phase obtained in the simulation, shown in Fig.\:\ref{f4}(b), consists of 4 NN lengths of about $2.54\ \mathring{\mbox{A}}$ and 4 NN lengths of about $2.69\ \mathring{\mbox{A}}$. These deviations of $\pm$\,3\% of the perfect B2 value cause the slight differences in the radial distribution function (Fig.\:\ref{f3}), but since they occur in an alternating manner, the simulation box is nevertheless cubic and the structure is closely related to a perfect B2. If the system is cooled down again, the emerging structure below the transition temperature [Fig.\:\ref{f4}(c)] consists of the same, but resorted NN distances as in B19$'$ [Fig.\:\ref{f4}(a)], which explains the nearly identical peak structure of the radial distribution function but a change in the shape of the simulation box. This structure has a cohesive energy of $E_c$\,=\,$5.078$ eV/atom, which is only about 0.04\% higher than $E_c\left(\mbox{B19}'\right)$, and it will be denoted as $\overline{\mbox{B19}'}$ in the following. Therefore it can be stated, that B19$'$ is very close to the global cohesive energy maximum within the used model. It is the authors opinion, that the small deviations of structures and lattice constants from experimental results and \emph{ab initio} calculations have to be traced back to the TBSMA potential model, which is an approximation and simplification of the real force field in NiTi.
\\
\indent In contrast to experiments, where a strong twinning of different martensitic variants is observed in NiTi nanocrystals when cooled down below the martensitic transition temperature \cite{waispi:05}, the ground state of the simulated system ($\overline{\mbox{B19}'}$) consists of one single variant. This may be due to the periodic boundary conditions and the simulation box as well as a system size, which is much smaller ($\approx 3-4\ \mathring{\mbox{A}}$) compared to the nanocrystals ($\approx 50\ \mathring{\mbox{A}}$).\\
\indent Since all the simulations are performed in a box with 2048 particles and periodic boundary conditions, it has to be ensured, that finite size effects do not play a decisive role concerning the structures, energetics and transition temperatures. To this end, the system size was varied between 500 and 32000 particles, whereupon the only differences were observed in shifts of the transition temperatures $T_A$ (B19$'$$\rightarrow$B2) of about 5~K and $T_M$ (B2$\rightarrow$$\overline{\mbox{B19}'}$) of about 10~K when going to the larger systems (see Fig.\:\ref{f10}). Furthermore, the involved structures and structural energies do not show a dependence on system size. The hysteresis between austenitic and martensitic transition takes a constant value when the system size exceeds 7000 particles, $T_A-T_M$\,=\,$35\pm 5$ K, in good agreement with recent experimental values \cite{khaami:09} (see also Sec.\:\ref{sec3B}). Therefore the existence of free energy barriers and associated hysteresis between low- and high-temperature structures is not an effect of finite system size, but rather a principal attribute of first-order martensitic phase transitions (see Sec.\:\ref{sec3C}). So it is reasonable to compare the transition temperatures resulting from the simulations with experimental values in the following.
\\

\subsection{Dependence on concentration}\label{sec3B}

\begin{figure}[!htbp]
\centering
\includegraphics[width=0.5\linewidth]{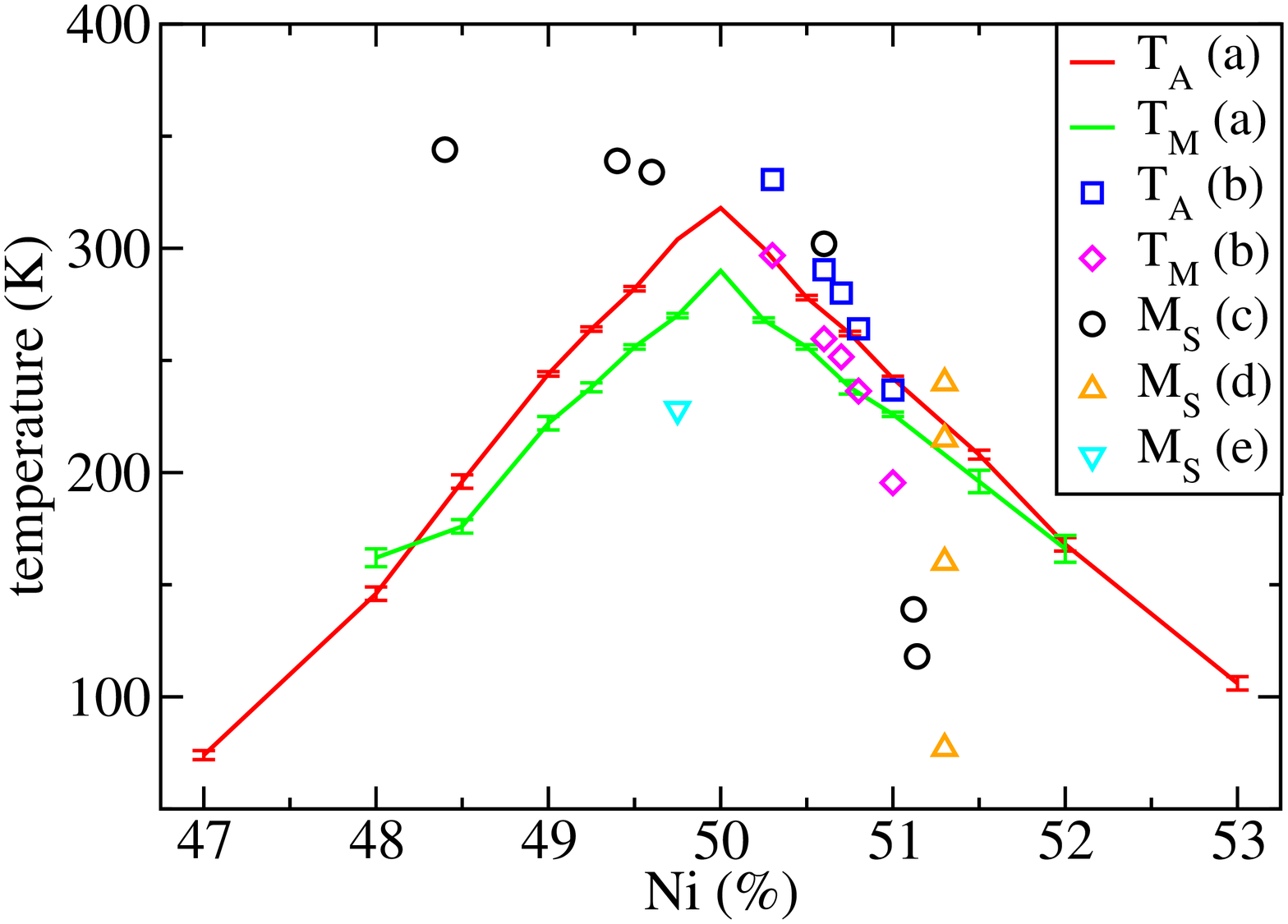}
\caption{Dependence of the transition temperatures on Ni concentration. Simulated data (a) in comparison with experimental values (b) \cite{khaami:09}, (c) \cite{hanbut:67}, (d) \cite{sabtat:82}, (e) \cite{otssaw:71}.}
\label{f5}
\end{figure}

\begin{figure}[!htbp]
\centering
\includegraphics[width=0.5\linewidth]{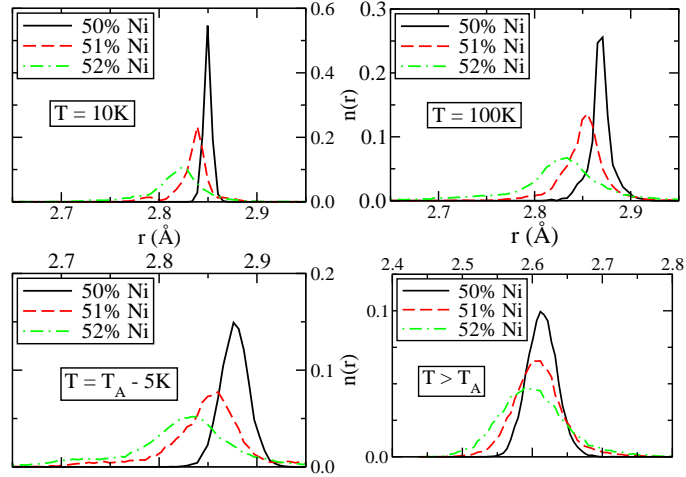}
\caption{Local structure analysis via the number of atoms with $d$\,=\,$r$ divided by the number of all atoms, $n\left(r\right)$, for 50\%, 51\% and 52\% Ni, at 3 temperatures below $T_A$ and one greater than $T_A$.}
\label{f6}
\end{figure}

\begin{figure}[!htbp]
\centering
\includegraphics[width=0.5\linewidth]{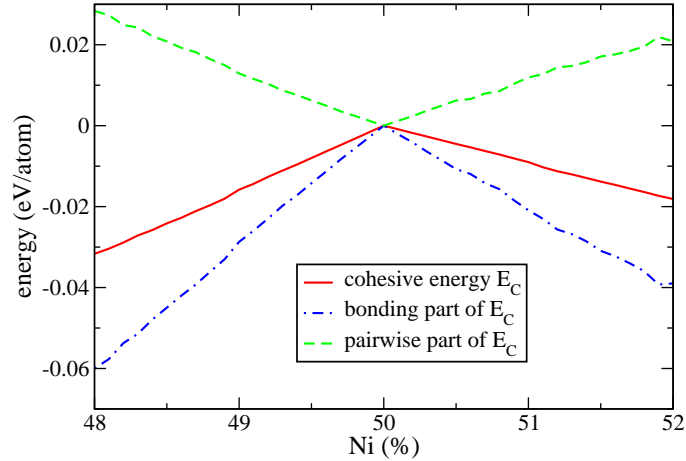}
\caption{Dependence of the equilibrium cohesive energy $E_c$, bonding part and pairwise interaction part of $E_c$ on Ni concentration. Energies are shown relative to the system with 50\ \% Ni.}
\label{f9}
\end{figure}

It is a well-known experimental fact, that temperatures, at which the structural changes between martensite and austenite in NiTi take place, are very sensitive to small deviations of the Ni/Ti concentration from the ideal one with 50\% Ni and 50\% Ti \cite{wasbut:71, hanbut:67, sabtat:82, khaami:09}. The effect, that B19$'$$\rightarrow$B2 transition temperatures in NiTi decrease with increasing Ni- or Ti-concentration is reasonable in a sense, since the crystallographic ground states of the pure materials Ni (FCC) and Ti (HCP) differ from a monoclinic structure corresponding to the alloy phase B19$'$. If the concentration of one of the constituents is rised, more and greater islands of pure material emerge, which are fixed in a matrix of an energetically less favored structure.\\
\indent In order to confirm this behavior theoretically within the semi-empirical approach, simulations were carried out where a temperature cycle was applied to systems with 2048 particles (as in Sec.\:\ref{sec3A}) and different Ni concentrations between 47\% and 53\%. Starting from a perfect Ni$_{50}$Ti$_{50}$ in the B19$'$ structure, these compositions were achieved by substituting a commensurate amount of atoms of one sort by atoms of the other sort randomly. In the heating process, the B19$'$$\rightarrow$B2 transition occurs instantaneously at a temperature denoted by $T_A$, and upon cooling, the system transforms into the $\overline{\mbox{B19}'}$ structure at the temperature $T_M$. For each concentration, 10 differently assembled systems were simulated. The averaged results are shown in Fig.\:\ref{f5}, together with experimental results for comparison. In the experiments, a difference is observed between the temperatures where the martensite nucleation starts ($M_s$) and ends ($M_f$), and where the austenitic transition begins ($A_s$) and ends ($A_f$), which leads to the definitions of $T_M$\,=\,$0.5\left(M_s+M_f\right)$ and $T_A$\,=\,$0.5\left(A_s+A_f\right)$.
\\
\indent On the Ni-rich side, transition temperatures agree well with recent results of Khalil-Allafi and Amin-Ahmadi \cite{khaami:09}. Since they measured differences $M_s-M_f$ and $A_f-A_s$ in the range of 29~K to 45~K, the simulated curves lie nearly in between these values. In addition, the extent of hysteresis, $T_A-T_M$\,$\approx$\,$30$~K, coincides well. For Ni concentrations $c_{\mbox{\tiny Ni}}$\,$\ge$\,$51\%$, the measured decrease of $M_s$ seems to be yet steeper than in the simulations, which is confirmed by the experiments of Hanlon \emph{et al.} \cite{hanbut:67}. It was pointed out by several authors \cite{tansun:99, prokor:04, otssaw:71, sabtat:82}, that experimental transition temperatures in NiTi depend strongly on the processes of material preparation before the actual measurement, since varying ageing temperatures and durations as well as quenching rates can lead to precipitation effects of intermediate phases. Experiments of Saburi \emph{et al.} \cite{sabtat:82} for an alloy with $c_{\mbox{\tiny Ni}}$\,=\,$51.3\%$ show for example $M_s$ values between 240~K and 77~K, depending on heat-treatment.\\
\indent The discrepancies between the simulated transition temperatures and the measured values at the Ti-rich side, where only very few experiments exist in the literature, are possibly due to Ti-precipitation effects in the measured samples, since lattice sites, where Ni atoms have been replaced by titanium, are attractive to each other \cite{luhu:07}. Therefore, they can form larger Ti islands influencing the transition behavior and temperatures. In the simulations, the replaced Ti or Ni atoms stay at their positions, leading to nearly symmetric TT curves around 50\% nickel.\\
\indent In order to explore the reasons for the strong dependence of $T_A$ and $T_M$ on $c_{\mbox{\tiny Ni}}$, the local environment of the atoms is examined. As shown in Fig.\:\ref{f4}(a) and (b) (see Sec.\:\ref{sec3A}), a structural feature for distinguishing unambiguously between B19$'$ and B2 locally is the mean length of the two nearest neighbor distances in $\left[\bar{1}01\right]$ and $\left[10\bar{1}\right]$ direction, since the elongations due to the monoclinic shear in B19$'$ vanish when the system transforms to B2. Defining this length as $d$ leads to $d\left(\mbox{B19}'\right)$\,=\,$2.85\ \mathring{\mbox{A}}$ at $T$\,=\,$0$~K and $d\left(\mbox{B2}\right)$\,=\,$2.61\ \mathring{\mbox{A}}$ at $T$\,$>$\,$T_A$ in Ni$_{50}$Ti$_{50}$ within the used potential model. Information about the local structures of all atoms in the system can be gained by evaluation of $n\left(r\right)$, the number of atoms with $d$\,=\,$r$ divided by the number of all atoms. This quantity is shown in Fig.\:\ref{f6} at four different temperatures during a heating process with starting configuration B19$'$, for systems with Ni concentrations of 50\%, 51\% and 52\%.
\\
\indent At $T$\,=\,$10$~K, the system with 50\% Ni shows a sharp peak at $r$\,=\,$2.85\ \mathring{\mbox{A}}$, as expected, whereas one additional percent of Ni lowers this height by a factor of about 0.5, what is proceeded when going to 52\% of Ni. Because of a constant number of particles, a lowering in height is equivalent to a broadening of the peaks, or more and more lengths $d$ differing from the ideal one at $T$\,=\,$0$~K. This is located at the peak maximum, which itself is shifted by about 0.2 $\mathring{\mbox{A}}$ per Ni-percent, resulting in smaller monoclinic shear angles $\alpha$.  Heating the system leads to thermal peak broadening and small horizontal shift, but the effect of concentration to the relative peak structure remains unchanged. So it can be stated, that a small deviation from $c_{\mbox{\tiny Ni}}$\,=\,$50\%$ is attended by a strong destabilization of the perfect B19$'$ lattice structure, independent of temperature, which can be regarded as a reason for the strong decrease of $T_A$ with $c_{\mbox{\tiny Ni}}$. At $T$\,=\,$T_A$, all atoms of the system transform collectively to B2 without any preceding structural changes, since the peaks are still located around $d$\,$\approx$\,$2.85\ \mathring{\mbox{A}}$ at temperatures slightly below $T_A$, but shifted completely to $d$\,$\approx$\,$2.61\ \mathring{\mbox{A}}$ at $T$\,$>$\,$T_A$, after the phase transition.\\
\indent Further insight into this destabilization effect can be gained by calculating the equilibrium cohesive energies for different Ni percentages between 48\% and 52\%. For this purpose, the system was set up in B19$'$ structure and the temperature was reduced stepwise to 0 K. The results are shown in Fig.\:\ref{f9}, together with the contributions of bond energy and pairwise interaction (Eq.\:\ref{g0}). The energy references are set to the values of the perfect structure with 50\% Ni. Varying the Ni percentage results in reduction of the cohesive energy in a symmetric way, which is strongly related to the decrease of transition temperatures (Fig.\:\ref{f5}). This behavior is due to the bonding part of the energy, which explains more instable lattices at Ni concentrations varying from 50\%.
\\
\indent Considering a perfect structure (p) on the one hand, and a structure with an ``impurity'' (i), where one Ti atom is replaced by a Ni atom on the other hand, cohesive energy differences $\Delta E_c = E_c^p-E_c^i$ of individual atoms at and around the impurity can be calculated. For the impurity atom itself, $\Delta E_c$\,=\,$0.523$~eV is obtained, as well as $\Delta E_c$\,=\,$0.026$~eV, if the impurity is a nearest neighbor atom, and $\Delta E_c$\,=\,$0.006$~eV in the case of a second nearest neighbor. This shows, that there is a lower value of $E_c$ in the direct vicinity of an additional Ni atom as in the surrounding perfect B19$'$ NiTi, which destabilizes the structure locally.\\
\indent Further inspection of the bond energy, according to the tight-binding parametrization (see Eq.\:\ref{g0}), reveals, that by replacing a Ti atom by a Ni atom, the parameters $q$, $\xi$, and $d$ change in such a way, that the bonding part of the cohesive energy of this Ni-``impurity'' is lower than for the Ti atom. Thus, the ``impurity'' scenario is energetically less favorable, resulting in an effective ``repulsion'' compared to the impurity-free case, destabilizing the lattice. Similar arguments apply for the Ti-rich side of the phase diagram, which explains the decrease of transition temperatures in this part of the phase diagram.

\subsection{Thermodynamics}\label{sec3C}

\begin{figure}[!htbp]
\centering
\includegraphics[width=0.5\linewidth]{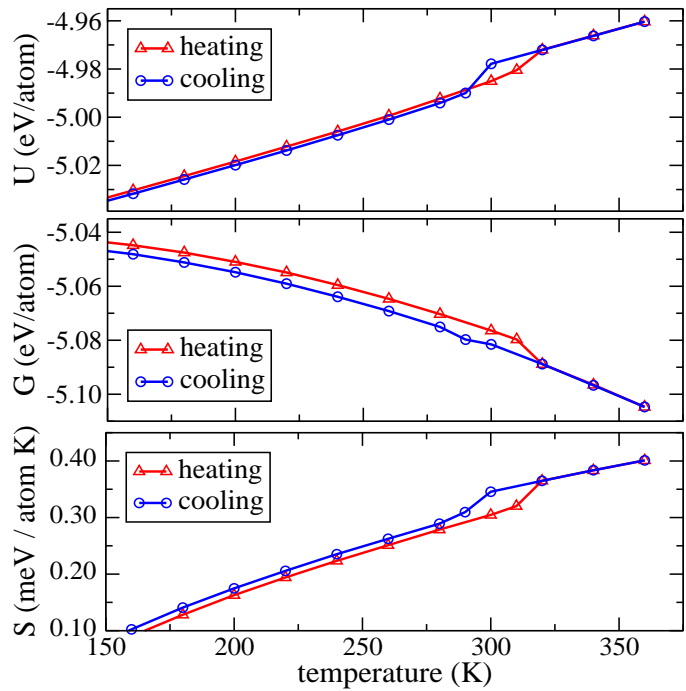}
\caption{Energy $U$, Gibbs free energy $G$ and entropy $S$ between $T$\,=\,$150$~K and $T$\,=\,$360$~K during a temperature cycle.}
\label{f7}
\end{figure}

\begin{figure}[!htbp]
\centering
\includegraphics[width=0.5\linewidth]{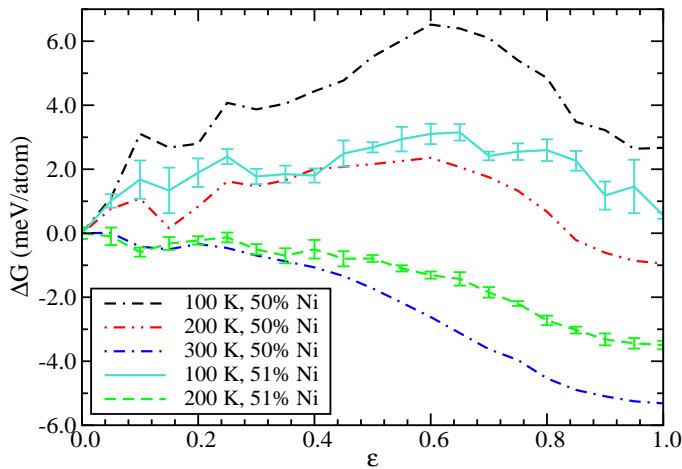}
\caption{$\Delta G_{T,c_{\mbox{\tiny Ni}}}\left(\epsilon\right) = G_{T,c_{\mbox{\tiny Ni}}}\left(\epsilon\right) - G_{T,c_{\mbox{\tiny Ni}}}\left(0\right)$ along a linear transformation path from B19$'$ ($\epsilon$\,=\,$0$) to B2 ($\epsilon$\,=\,$1$) for temperatures $T$\,=\,$100$~K, $200$~K, $300$~K and nickel concentrations $c_{\mbox{\tiny Ni}}$\,=\,$50\%$, $51\%$. The error bars at the values for $c_{\mbox{\tiny Ni}}$\,=\,$51\%$ result from averaging over 3 randomly assembled systems.}
\label{f8}
\end{figure}

Martensite/austenite phase transitions can in general be explained by means of different competing contributions to $G$, the Gibbs free energy. One phase can only form within a matrix of the other one, if non-chemical elastic strain energies and energy differences due to unlike interfaces ($\Delta G_{nc}$) are overcome by a chemical energy $\Delta G_c$ \cite{buchots:98}. Therefore, the structural transition occurs, if $\Delta G_c$\,=\,$\Delta G_{nc}$, and leads to a jump of $G$ about this amount, which characterizes the transition as first-order. Thermodynamic properties of the phase transitions visible in the presented simulations are calculated by thermodynamic integration as described in Sec.\:\ref{sec2B}. A 2048 particle system in B19$'$ configuration is heated above $T_A$ to $T$\,=\,$360$~K and subsequently cooled. At temperature intervals of 20~K (10~K in the vicinity of the structural change), the Gibbs free energy is determined. Together with the internal energy $U$ and the entropy $S=\left(U-G\right)/T$, $G$ is shown in Fig.\:\ref{f7} between $T$\,=\,$150$~K and $T$\,=\,$360$~K.
\\
\indent At $T$\,=\,$T_A$, $G$ drops about $\Delta G_c$\,$\approx$\,$9\ \mbox{meV/atom}$ from $G_{\mbox{\scriptsize B19}'}$ to the lower $G_{\mbox{\scriptsize B2}}$. Since at this temperature $U$ increases stronger than only due to thermal effects as in the previous heating, the negative $\Delta G$ has to be accompanied by a jump in entropy to a higher value. Khalil-Allafi and Amin-Ahmadi measured austenitic start- and end-temperatures ($A_s,A_f$) as well as enthalpies $\Delta H_{M\rightarrow A}$ in B19$'$$\rightarrow$B2 transitions for NiTi with Ni content between 50.3\% and 51.0\% \cite{khaami:09}. $T_A$\,=\,$0.5\left(A_s+A_f\right)$ and $\Delta H_{M\rightarrow A}$ show nearly linear behavior with composition in this range and can therefore be extrapolated to 50\% Ni, what leads to $\Delta S_{M\rightarrow A}$\,=\,$\Delta H_{M\rightarrow A}/T_A$\,=\,0.044~$\mbox{meV/(atom$\cdot$K)}$, in good agreement with the value obtained in the simulation of 0.045~$\mbox{meV/(atom$\cdot$K)}$. Several \emph{ab initio} studies \cite{huabun:01, parpar:02, hatkon:09.1, souleg:10} state an entropic stabilization of the high temperature phases in NiTi by vibrational entropy and phonon mode softening, respectively. If, in the present simulation, the temperature is decreased again, a jump of $G$ at $T$\,=\,$T_M$ back to the B19$'$-curve is not observed, because the system does not transform to B19$'$ upon cooling, as explained in Sec.\:\ref{sec3A}. Nevertheless, investigation of $U$ and $S$ shows the occurence of a phase transition (B2$\rightarrow$$\overline{\mbox{B19}'}$) by revealing the contrary behavior to the heating process: the energetically more favorable structure with lower entropy is adopted by the system.\\
\indent Moreover, with the help of free energy calculations it is possible to detect free energy barriers in transformation paths between the occurring structures, which are typical in first-order phase transitions and responsible for hysteresis. As an example, a linear path between B19$'$ and B2 is modeled, and at each intermediate step, the free energy is calculated by a simulation as above. To this end, the system is set up in a structure between B19$'$ and B2 with edge lengths $a,\ b,\ c$ and an angle $\alpha$ of the unit cells [see Fig.\:\ref{f1}(a)] according to:
\begin{eqnarray}
a\left(\epsilon\right) &=& a_{\mbox{\scriptsize B19}'} + \epsilon\cdot\left(a_{\mbox{\scriptsize B2}}-a_{\mbox{\scriptsize B19}'}\right)\nonumber\\
b\left(\epsilon\right) &=& b_{\mbox{\scriptsize B19}'} + \epsilon\cdot\left(b_{\mbox{\scriptsize B2}}-b_{\mbox{\scriptsize B19}'}\right)\nonumber\\
c\left(\epsilon\right) &=& c_{\mbox{\scriptsize B19}'} + \epsilon\cdot\left(c_{\mbox{\scriptsize B2}}-c_{\mbox{\scriptsize B19}'}\right)\nonumber\\
\alpha\left(\epsilon\right) &=& \alpha_{\mbox{\scriptsize B19}'} + \epsilon\cdot\left(\alpha_{\mbox{\scriptsize B2}}-\alpha_{\mbox{\scriptsize B19}'}\right)\nonumber
\end{eqnarray}
with a parameter $\epsilon$ varying from 0 (B19$'$) to 1 (B2), and values for the reference structures obtained in this work (see Tab.\:\ref{tab1}). The angles $\beta$ (between $a$- and $b$-edge) and $\gamma$ (between $b$- and $c$-edge) remain at 90$^{\circ}$. The system is forced to stay in this structure by a stiff simulation box with edge lengths and angles according to the unit cells values, but the larger the system size, the less is a stiff box able to hold the system into a defined structure. Therefore the number of particles is reduced to 500 in this study. Fig.\:\ref{f8} shows the results at different temperatures $T$ and Ni concentrations $c_{\mbox{\tiny Ni}}$ in form of $\Delta G_{T,c_{\mbox{\tiny Ni}}}\left(\epsilon\right) = G_{T,c_{\mbox{\tiny Ni}}}\left(\epsilon\right) - G_{T,c_{\mbox{\tiny Ni}}}\left(0\right)$. Thus, the differences in $G$ can be seen along the transformation path and compared for varying $T$ and/or $c_{\mbox{\tiny Ni}}$, without accounting for the absolute $G$ values.
\\
\indent Regarding the system with $c_{\mbox{\tiny Ni}}$\,=\,$50\%$, a free energy barrier can be detected for $T$\,=\,$100$~K, which flattens when going to higher temperatures. The B2 structure lies in a minimum of $\Delta G$, too, but this is not surprisingly, since the potential parameters were determined by fitting to material properties of B2 at $T$\,=\,$0$~K. Nevertheless, at low temperatures, B19$'$ exhibits a lower $\Delta G$, but already at $T$\,=\,$200$~K, more than 100~K below the phase transition, B2 would be more stable, and only the barrier in $\Delta G$ prevents the system from transforming. Upon cooling, the barrier does not disappear, which explains, why a B2$\rightarrow$B19$'$ transition is not observed in the simulations. Increasing the Ni content to 51\% leads to a less pronounced free energy barrier at $T$\,=\,$100$~K with a more kinked curve shape, which is caused by slight reorientations of the initially set up structures, and to considerable uncertainties when averaging over 3 differently assembled systems. This behavior reflects the result of Sec.\:\ref{sec3B}, where a more unstable and thus more fluctuating lattice structure has been obtained by varying the Ni concentration away from 50\%. As a consequence, transition temperatures decrease, what is confirmed here, too, since the $\Delta G$ barrier vanishes already at about 200~K.

\section{Summary and conclusion}\label{sec4}

In this work, MD simulations of the temperature driven structural phase transitions in NiTi alloys with equi- or nearly equiatomic composition were carried out by using a semi-empirical model potential from the literature \cite{lailiu:00} (TBSMA), since this is known to predict a monoclinic structure to be energetically more favorable than a cubic one at $T$\,=\,$0$~K.\\
\indent It could be shown, that a B19$'$ structure is stable within this model at low temperatures, and a slight modification concerning the cutoff behavior of involved functions leads to good agreement of lattice parameters and energetics with \emph{ab initio} and experimental results. By analyzing simulation box shape, radial distribution function (RDF) and nearest neighbor (NN) environments during an increase followed by a decrease of the imposed temperature, structural phase transitions were observed. While at $T$\,=\,$T_A$ during heating (318~K for Ni$_{50}$Ti$_{50}$) the system adopts a nearly cubic structure closely related to B2, at a temperature $T$\,=\,$T_M$ upon cooling, a structure denoted $\overline{\mbox{B19}'}$ emerges. This has nearly the same RDF than B19$'$, but the NN's belonging to the RDF peaks are resorted, resulting in a larger cohesive energy of about 0.04\%. So it can be stated, that the TBSMA approach predicts a stable B19$'$, which is very close to the martensitic ground state $\overline{\mbox{B19}'}$ of this model.\\
\indent The experimentally known fact, that transition temperatures (TT) vary strongly with nickel concentration was confirmed qualitatively at the Ni-rich side and even quantitatively in the range between 50\% and 51\% nickel. At the Ti-rich side, there exist only a few and partly conflicting experimental results, which do not fit well with the simulated curves. The discrepancies may be due to precipitation effects in the material processing before the measurements, which do not emerge in the simulations. By investigation of NN distances for systems with $c_{\mbox{\tiny Ni}}$\,$\ge$\,$50\%$ during heating, it could be shown, that the strong decrease of the austenitic TT is attended by a destabilization of the B19$'$ lattice structure. This destabilization results from the bonding part of the cohesive energies in the direct vicinity of a Ni-``impurity'', which is lower than in a perfect B19$'$ NiTi. A relation between lattice stability and TT values was also proposed by Lu \emph{et al.} \cite{luhu:07}, who performed \emph{ab initio} charge density calculations of B2 Ni-rich NiTi.\\
\indent Thermodynamic calculations gave more insight into the phase transition process by confirming that the high temperature phase is entropically stabilized, with a jump of $\Delta S$ at $T$\,=\,$T_A$ fitting well with recent experimental results \cite{khaami:09}. Free energy barriers suppressing the phase transitions were determined along a linear path between B19$'$ and B2, and it could be shown, that the heights of these barriers are lowered by increasing on the one hand temperature and on the other hand nickel concentration from 50\% to 51\%.

\begin{acknowledgments}
We gratefully acknowledge the support of the SFB 513, the SFB 767, the NIC, and the HLRS.
\end{acknowledgments}

\end{document}